\documentclass[final]{aipproc}

\layoutstyle{6x9}

\usepackage{epsfig}

\newcommand{\as}{\bar\alpha_s}

\newcommand{\eq}[1]{(\ref{#1})}
\newcommand{\wid}{0.4\columnwidth}

\begin{document}

\title{Saturation, traveling waves and fluctuations\footnote{Talk presented at
the DIS 2005 workshop, April 2005, Madison, Wisconsin, and at the XIth International Conference on Elastic and Diffractive Scattering, May 2005, Blois, France.}}


\author{Rikard Enberg}{
address={Theoretical Physics Group, Lawrence Berkeley National Laboratory, CA 94720, USA},
address={Centre de Physique Th{\'e}orique,
{\'E}cole Polytechnique, 91128~Palaiseau, France}
}

\begin{abstract}
In this talk I discuss the high energy asymptotics of QCD scattering, and its similarity to a reaction--diffusion process. I also discuss detailed numerical studies of the mean field approximation to this picture, i.e., the Balitsky--Kovchegov equation.
\end{abstract}

\maketitle

Perturbative QCD, in the form of the BFKL equation, predicts that in the high energy (small-$x$) limit, cross sections rise as a power of the center of mass energy. This rise is not compatible with unitarity of the $S$-matrix when extrapolated to very high energies. The BFKL equation is a linear evolution equation; the limits set by unitarity requires the introduction of non-linear terms in the evolution equation. The simplest such evolution equation is the Balitsky--Kovchegov (BK) equation \cite{BK}. 

More involved QCD evolution equations exist; they are known as the Balitsky--JIMWLK equations. Very recently, it has been discovered that there are important effects that are not accounted for by this set of equations (see \cite{StastoKovner} and references therein). 

A different point of view has recently been emerging, in which high energy QCD is equivalent to a reaction--diffusion system. First the BK equation was shown \cite{MP2003} to be in the equivalence class of the Fisher--Kolmogorov--Petrovsky--Piscounov (FKPP)  non-linear partial differential equation, which has so-called traveling wave solutions. This
allowed deriving results for the amplitude and saturation scale as a function of energy. Then it was realized \cite{M2005} that the effects neglected in the BK equation can be described within the equivalence class of the \emph{stochastic} version of the FKPP equation. 

This talk is based on our paper \cite{EGM} where we wanted first to understand in more detail how the statistical interpretation of high-energy QCD comes about and then to make detailed numerical studies of the properties of the mean field approximation (the BK equation) and of a toy model that captures the important features of high energy QCD. 

Let us first discuss the mean field approximation of the full evolution, which yields the BK equation.
The first equation in Balitsky's hierarchy of coupled equations can be written as
\begin{equation}
\partial_{\as Y} \langle T(k,Y) \rangle =
K\otimes \langle T(k,Y) \rangle -
 \langle T^2(k,Y) \rangle,
\end{equation}
where $L=\ln(k^2/\Lambda^2)$, and $K \otimes $ means the action of the BFKL integral kernel. The next equation in the hierarchy is an equation for the evolution of the correlator $\langle T^2(k,Y) \rangle$. If one approximates this correlator as the factorized form $\langle T^2(k,Y) \rangle \approx \langle T(k,Y) \rangle^2 \equiv A^2(k,Y)$ one gets the BK equation for the mean-field amplitude $A$:
\begin{equation}
\partial_{\as Y} A(k,Y) = K\otimes A(k,Y) - A^2(k,Y) 
\end{equation}
Munier and Peschanski showed that this equation is in the equivalence class of the FKPP equation,
$
\partial_t u(x,t) = \partial^2_x u(x,t) + u(x,t) - u^2(x,t),
$
where $x$ corresponds to $L$ and $t$ corresponds to $Y$. This equation has traveling wave solutions for certain conditions on the initial conditions, see \cite{MP2003}. This means that there is a solution which has a more or less fixed shape, and under the evolution in time (rapidity), the position of the wave front moves in space (momentum). Using results from the study of the FKPP equation, they obtained an expression for the saturation scale as a function of rapidity,
%
%
and analytical expressions for the shape of the wave front for momenta above the saturation scale. 

In \cite{EGM}, we solved the BK equation numerically and studied the properties of the solution. In particular we compared the numerical results to the analytical results. In Fig.\ 1, we see that there are indeed traveling wave solutions, and that the analytical results agree very well with the full solution for large rapidities.
We also confirm the prediction that the influence of the initial condition disappears for large $Y$, so that a universal propagation speed is approached.

\begin{figure}[th]
\epsfig{file=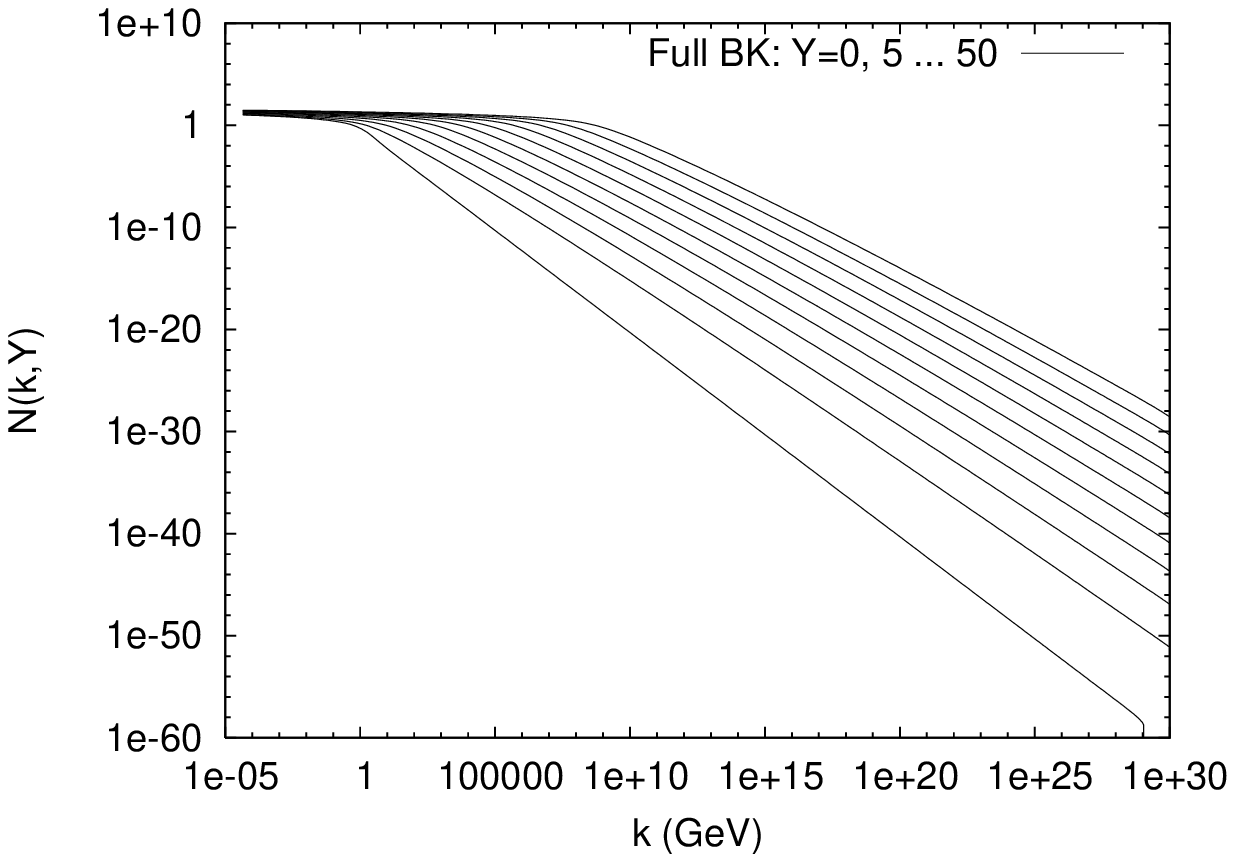,width=\wid}%
\epsfig{file=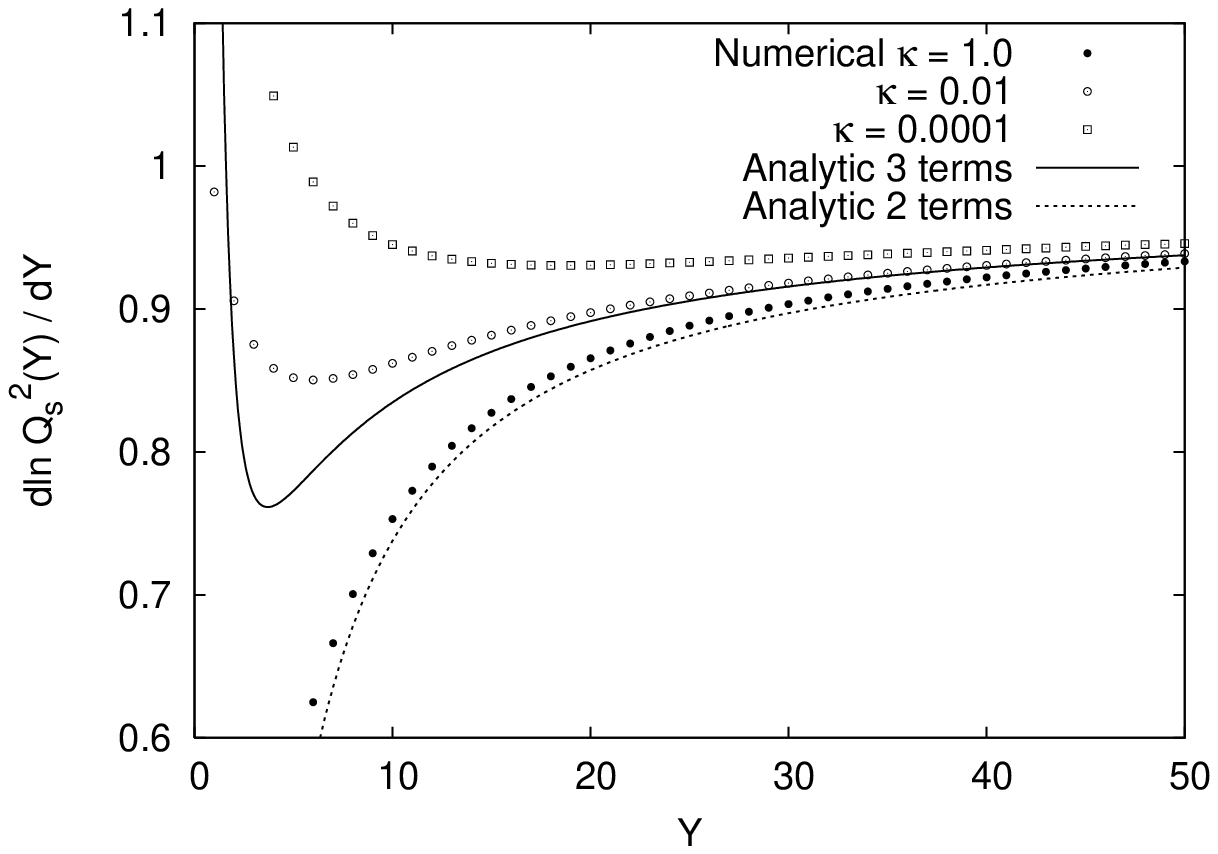,width=\wid}
\caption{Left: Evolution of the initial condition using the full 
BK equation. Right: Logarithmic derivative of the saturation scale 
${d \ln Q_s^2(Y)}/{d Y}$ obtained from numerical 
simulations, compared to the analytical results to two different levels of accuracy.
\label{fronts1}}
\end{figure}

The stochastic FKPP equation describes physical situations in which objects evolve by multiplying and diffusing, up to a limiting threshold.
The crucial point is that the QCD parton model has such dynamics  \cite{M2005,IMM,EGM}. Once this is realized, all mathematical results obtained in statistical physics can be transposed to high energy QCD \cite{IMM}.

In the dipole picture of high energy scattering, a QCD dipole scatters off a hadronic target by interacting with one of its quantum fluctuations. The probe effectively ``counts'' the partons
in this Fock state of the target
with transverse momenta $k$:
the amplitude $T(k)$
for the scattering is related to the number of partons $n(k)$.

The wave function of a hadronic object 
is built up from successive splittings
of partons starting from the valence
structure.
As one increases the rapidity $Y$,
the phase space for parton splittings grows and
makes the probability for high occupation numbers larger.
In the initial stages of the evolution, the parton
density grows diffusively from these splittings.
On the other hand, the number of partons in each cell of transverse 
phase space is limited 
to a maximum number $N$ that depends on 
the strength of the interaction of the probe,
i.e., on the relationship between $T$ and $n$. This property is necessary from
unitarity, which imposes an upper bound on the amplitude (e.g. $T(k)\leq 1$) which, in turn, results in an
upper bound on $n(k)$. 
This is parton saturation.

Viewed in this way, scattering in the 
parton model is a reaction--diffusion process. 
The rapidity evolution of the Fock states of the target hadron
is like the time evolution of a set of particles that diffuse in space, 
multiply and recombine
so that on average there are no more than $N$ particles
on each site. 
The QCD amplitude $T$ corresponds to the fractional occupation number
$u=n/N$ in the reaction--diffusion model. In the continuum limit, $u$ obeys the so-called Reggeon field theory equation (which is in the same equivalence class as the stochastic FKPP equation)
\begin{equation}
\partial_t u(x,t)=D\partial_x^2 u(x,t)+\lambda u(x,t)-\mu u^2(x,t)
+ \sqrt{(2/N)u(x,t)}\,\eta(x,t)\ ,
\label{RFT}
\end{equation}
where $\eta$ is a Gaussian white noise function. 
We take this as a model of QCD scattering at high energy. The model does not give the exact behavior of the scattering, but should describe its gross features in the saturation limit. The second order differential operator should be replaced with the BFKL kernel in real QCD. See \cite{EGM,IT} for a detailed discussion of the correspondence to QCD.

Eq.\ \eq{RFT} describes the scattering off one partonic realization of the target. The amplitude $u$ is a random variable, which fluctuates between different realizations of a scattering. To get the physical amplitude one must take the average of all realizations.

Taking into account the properties of the noise term  leads to a hierarchy of equations for correlators similar to the Balitsky hierarchy, which, however, has some extra boundary terms that are important (see \cite{EGM}). The mean field approximation leads, in this case, to the FKPP equation, in analogy with the BK case. 

To study this numerically, we construct a toy model with a number $N$ of particles on a one-dimensional lattice, that are allowed to jump to the neighboring sites, to multiply, and to disappear with certain probabilities. This type of model has been extensively studied in statistical physics \cite{Panja} and allows straightforward simulation. For the exact description of the model, see \cite{EGM}. Here, we just note that the number $N$ corresponds to $1/\alpha_s^2$ in QCD, so the many-particle limit corresponds to small $\alpha_s$.

\begin{figure}
\epsfig{file=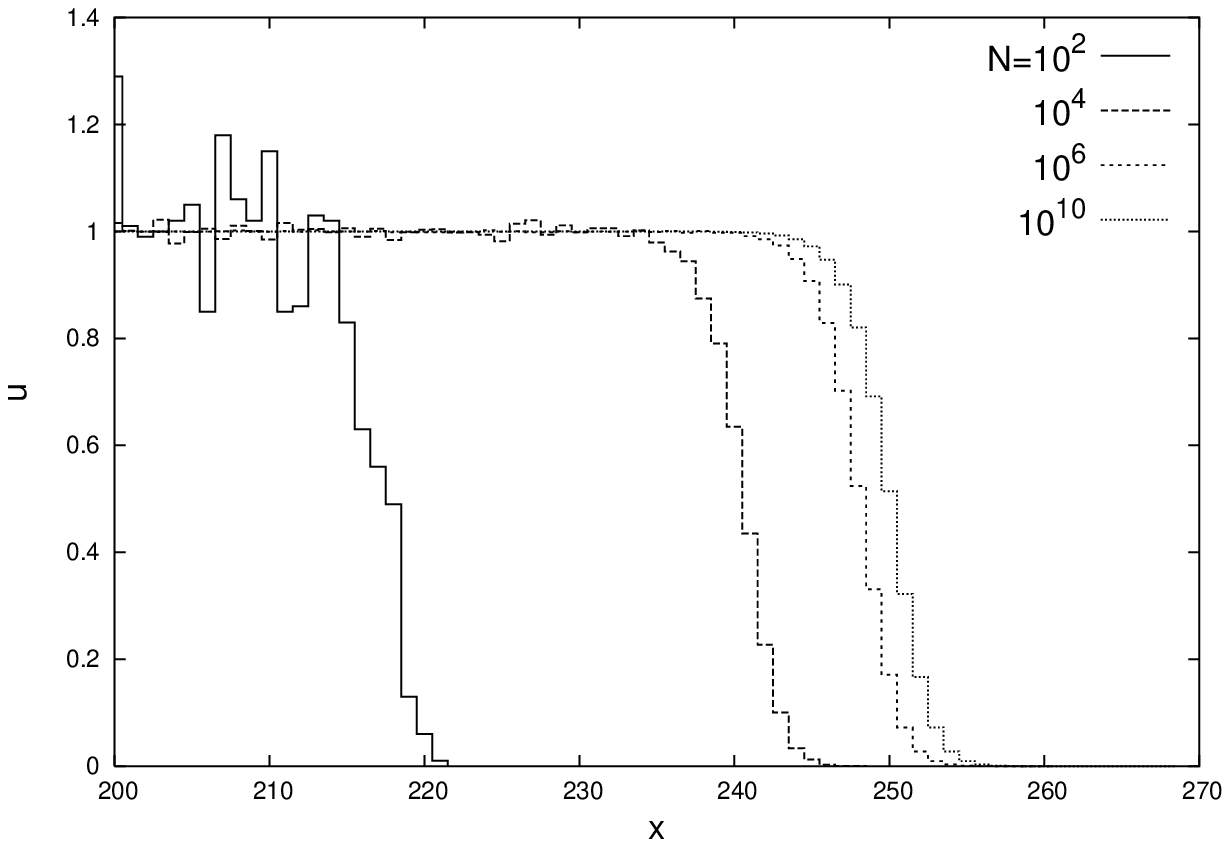,width=\wid}
\epsfig{file=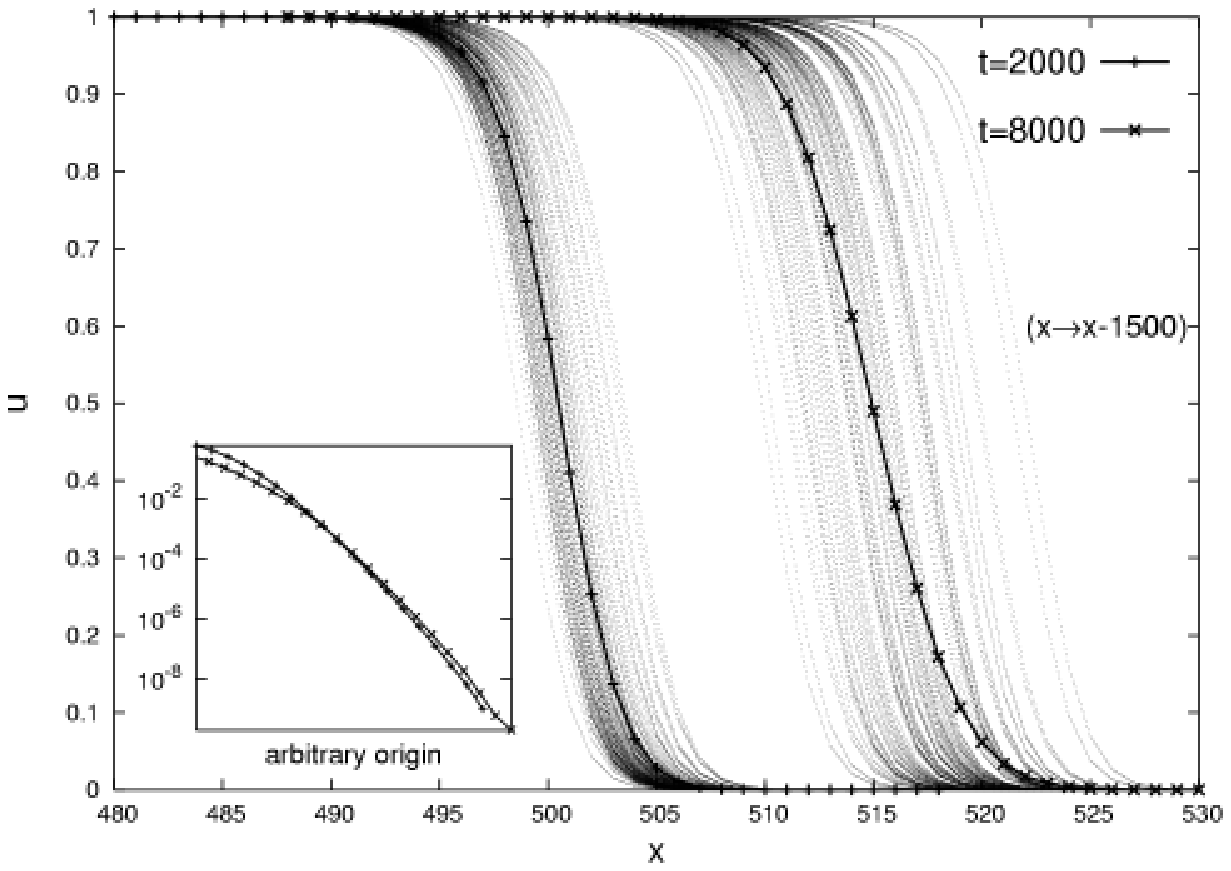,width=\wid}
\caption{\label{front1}Left: Numerical integration of the toy model over 1000 units of time for different values of $N$. Right: 1000 realizations of the evolution of the toy model
between time 0 and $t_1=2000$ (left bunch of curves), and $t_2=8000$
(right bunch of curve). {\it Insert:} the average front for these two times.}
\end{figure}

\begin{figure}
\epsfig{file=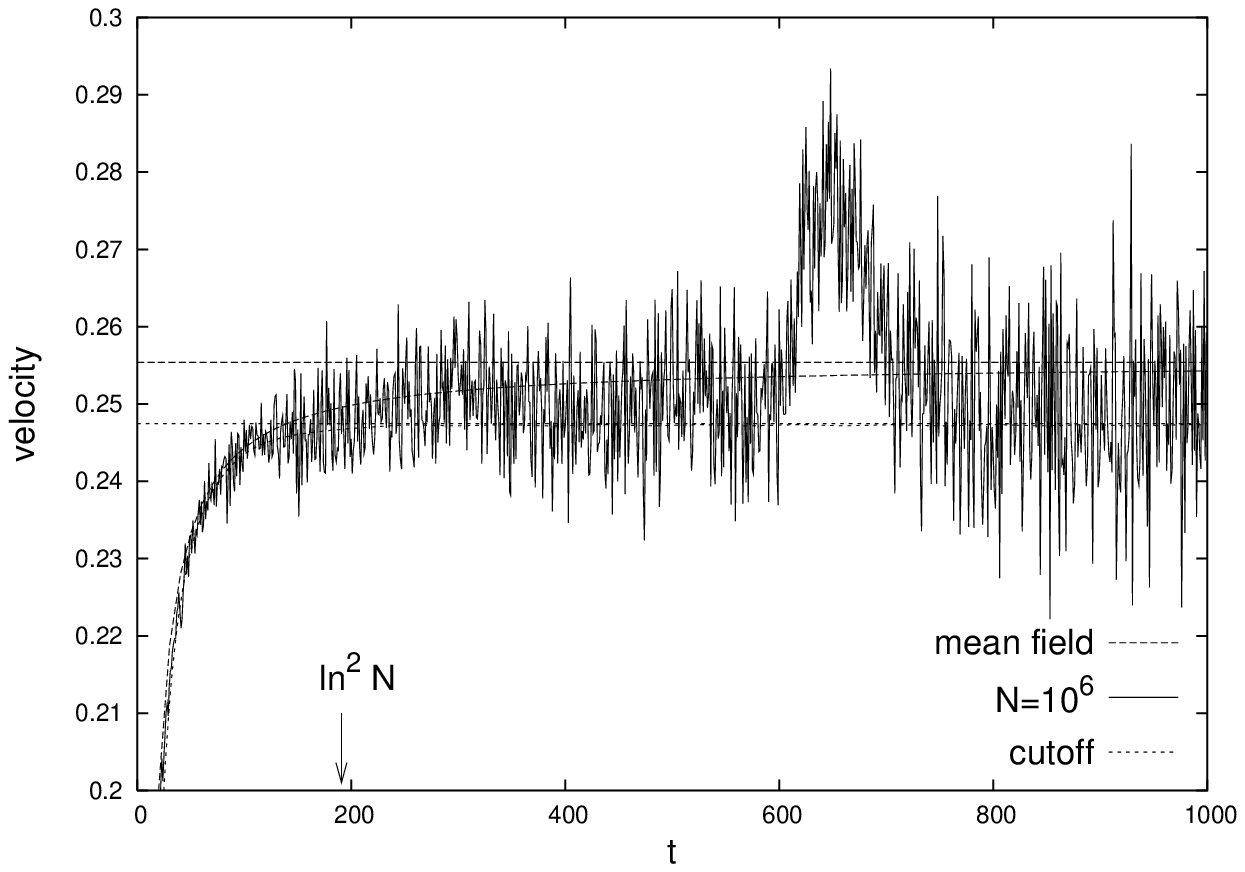,width=\wid}
\epsfig{file=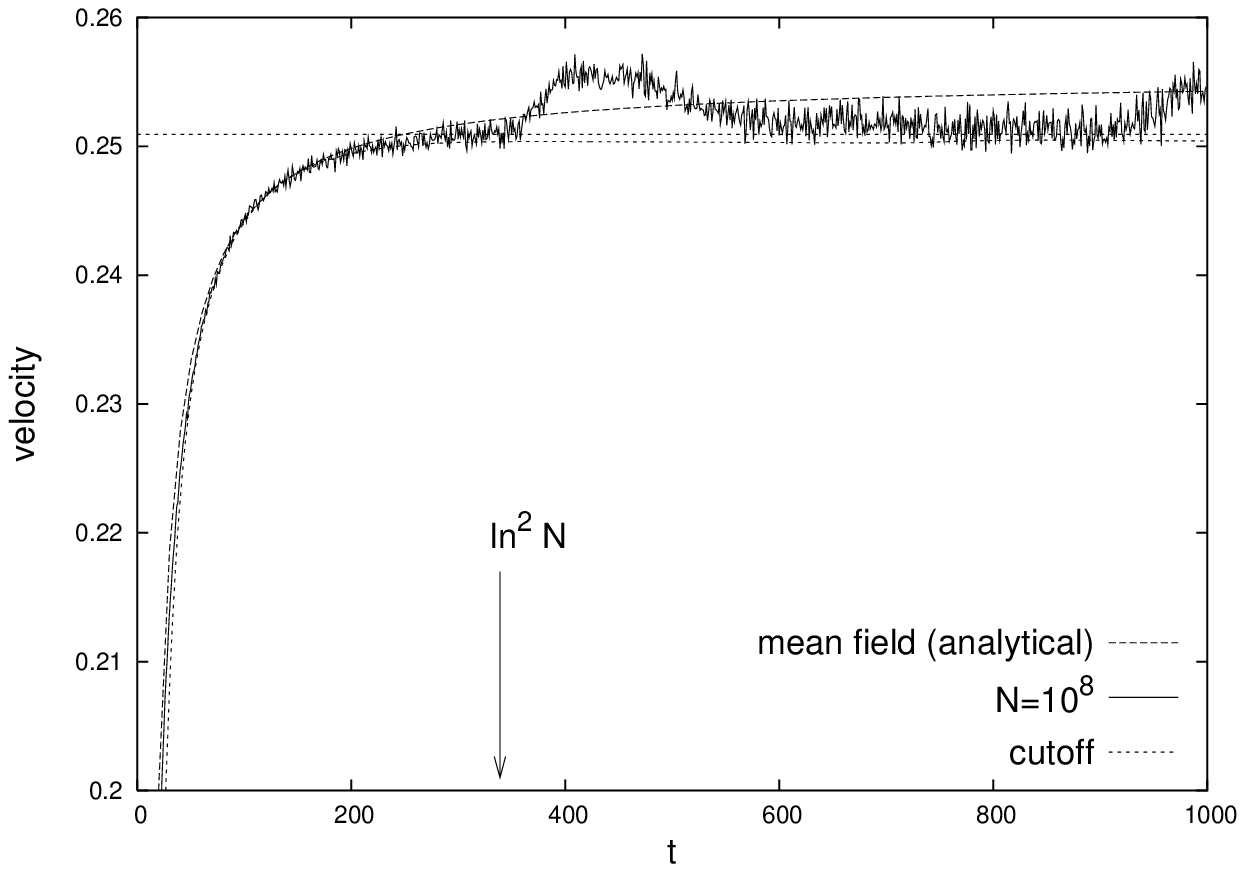,width=\wid}
\caption{Instantaneous velocity of the front as a function of time (full fluctuating line)
from a numerical solution of the stochastic evolution equation
for $N=10^6$ and $N=10^8$ particles per site.}
\end{figure}

The left plot in Fig.\ 2 shows the evolution of one realization of the numerical model for different $N$. The fronts have fluctuations around the steady front shape without fluctuations, with smaller fluctuations for larger $N$. 

The large time velocity of the wave front for a system described by the model \eq{RFT} has been shown to be \cite{BD}
$v=v_0-c/\ln^2 N,$
where $v_0$ is the velocity in the absence of fluctuations, and $c$ is a constant.
Fig.\ 3 shows the velocity of one realization as a function of time for two different $N$, showing appreciable fluctuations. The dashed curves show the analytical predictions; for large times the simulation gets closer to the asymptotic value; for smaller times the mean field prediction is still  good. This also means that geometric scaling is expected to hold for intermediate times (or rapidities in QCD).

The right plot in Fig.\ 2, finally, shows the breaking of geometric scaling that arises from the averaging. Since the position of the front at a fixed time fluctuates between different realizations, the average will have a different shape than each individual curve. This is illustrated in the plot at two different times. At the earlier time the breaking is not so large, but at the larger time it is much more pronounced, as is the spread in position.

Although the mean field approximation seems to remain useful for intermediate, phenomenologically accessible energies (see also \cite{Soyez}), it is important to understand the corrections to this picture.
The main interest of the statistical approach described above is to get a simple physical understanding of the fluctuations in the (extended) Balitsky-JIMWLK equations, and to make it possible to perform a simple derivation of the universal asymptotics of the scattering amplitude at high energy. The simplicity comes from the reaction--diffusion nature of QCD scattering at high energy. 

\emph{Acknowledgment}
This talk is based on work done in collaboration with Krzysztof Golec-Biernat and St\'ephane Munier \cite{EGM}.

\end{document}